%% file: FalseData_ColoredGaussian.tex
\documentclass[conference]{IEEEtran}

\usepackage{amsthm}
\usepackage{makecell}
\usepackage{tabularx}
\usepackage{setspace}

\usepackage{algpseudocode}
\usepackage[dvips]{graphicx}
\usepackage{array}
\usepackage{threeparttable}
\usepackage{amsmath}
\usepackage{multirow}
\usepackage{booktabs}
\usepackage{graphicx}
\usepackage{tabularx}
\usepackage{xcolor}
 \usepackage[caption=false,font=footnotesize]{subfig}%

\newcommand{\removelatexerror}{\let\@latex@error\@gobble}

\include{Macros}

\newcommand*{\Scale}[2][4]{\scalebox{#1}{$#2$}}%
\DeclareMathOperator*{\argmax}{arg\,max}

\begin{document}

\title{Detection of False Data Injection Attacks in Smart Grid under Colored Gaussian Noise}
\author{Bo Tang,\IEEEmembership{Student Member, IEEE,} Jun Yan,\IEEEmembership{Student Member, IEEE,} Steven Kay,\IEEEmembership{Fellow, IEEE,} and Haibo He\IEEEmembership{Senior Member, IEEE}\\ 
\IEEEauthorblockA{Department of Electrical, Computer, and Biomedical Engineering\\
University of Rhode Island\\
Kingston, Rhode Island 02881\\
Email: \{btang, jyan, kay, he\}@ele.uri.edu}
}
\maketitle


\begin{abstract}
In this paper, we consider the problems of state estimation and false data injection detection in smart grid when the measurements are corrupted by colored Gaussian noise. By modeling the noise with the autoregressive process, we estimate the state of the power transmission networks and develop a generalized likelihood ratio test (GLRT) detector for the detection of false data injection attacks. We show that the conventional approach with the assumption of Gaussian noise is a special case of the proposed method, and thus the new approach has more applicability. {The proposed detector is also tested on an independent component analysis (ICA) based unobservable false data attack scheme that utilizes similar assumptions of sample observation.} We evaluate the performance of the proposed state estimator and attack detector on the IEEE 30-bus power system with comparison to conventional Gaussian noise based detector. The superior performance of {both observable and unobservable false data attacks} demonstrates the effectiveness of the proposed approach and indicates a wide application on the power signal processing. 
\end{abstract}




\IEEEpeerreviewmaketitle

\section{Introduction}

The power system relies on accurate measurements of system topology and state variables to analyze real-time system dynamics and maintain stable operation. However, measurements collected by the supervisory control and data acquisition (SCADA) system are often corrupted by random noises or missing data \cite{sood2009developing}. To recover accurate system variables and detect potential bad data for system operation, robust state estimation techniques are commonly used in the energy management systems (EMS) \cite{monticelli2000electric}.

However, as the smart grid brings in cybernetic integration with the computerized communication network to the modern electrical power infrastructure, the industry and research community have witnessed growing security concerns from false data injection (FDI) attacks in state estimation \cite{sandberg2010security, hug2012vulnerability, giani2013smart, tan2015integrity, yu2015blind}. Due to the linear approximation in state estimation, malicious attackers can construct stealth schemes to inject random or targeted false data into the power system measurements that can result in serious instability in system operation. 

{
Although numerous studies have investigated the mathematical methods to build robust state estimation and detection mechanisms against the FDI attacks, they all are built on a common assumption that the background noise in state estimation are white Gaussian noise (WGN) \cite{Cui2012, liu2011false, teixeira2010cyber}. However, many natural phenomena, such as ice cracking and atmospheric noise, and man-made noise sources, such as electronic devices, can be modeled more accurately as non-Gaussian distributions \cite{plataniotis1997nonlinear}\cite{pitas2013nonlinear}. Detection performance of conventional state estimator and false data detector would deteriorate due to the presence of colored Gaussian noise or even more complex non-Gaussian noises.}

In this paper, we investigate the problems of state estimation and false data injection detection when the measurements are corrupted by the colored Gaussian noise. We model the colored Gaussian noise via the autoregressive (AR) process, and derive a closed form of state estimation and a generalized likelihood ratio test (GLRT) detector for false data injection detection. This paper has shown the deterioration of conventional WGN-based bad data detector in colored Gaussian noise and proposed an AR-based detector. Moreover, it can be shown that the conventional Gaussian noise approach is the special case of the proposed AR approach. The computer simulations on the IEEE 30-bus power system are conducted to evaluate the detection performance of the proposed AR approach. 

Throughout this paper, we use the boldfaced uppercase character (e.g., $\mathbf{X}$) to denote the matrix, the boldfaced lowercase character (e.g., $\mathbf{x}$) to denote the vector, and the unboldfaced character (e.g., $n$ or $N$) to denote the scalar. The symbols of $[\cdot]^{T}$ and $[\cdot]^{-1}$ denote the transpose and the inverse of a matrix, respectively.

\section{Background and Related Work}
State estimations are performed on power transmission networks which consist of a set of generators, load buses, transmission lines, and other electrical facilities governed by physical laws. Measurements are collected on meters, e.g., voltage meters and phasor measurement units (PMU), and reported to wide-area control centers through the SCADA system. The state variables of the power system is further estimated at the control centers from these measurements with the knowledge of power system topology. Control actions are then determined to maintain stable, cooperative transmission from power plants to customers in interconnected power grids.

Traditional state estimators are capable of identifying and eliminating bad data from state estimations \cite{xu2011state}\cite{hagh2011improving}. However, it has been recently found that malicious data attacks can exploit system topology information to construct false data injection schemes that will bypass bad data detectors in conventional state estimation \cite{anwar2014vulnerabilities}\cite{dan2010stealth}\cite{Liyan2015}. A number of defending strategies have been proposed \cite{vukovic2012network,huang2013bad,kim2013topology}. Studies also proposed optimal PMU placement to detect FDIs in the smart grid. Recently, machine learning algorithms have also been introduced to detect stealth phases \cite{esmalifalak2013detecting}. These studies, however, are based on the assumption of white Gaussian noise embedded in the measurements. Their performance remains unknown in the presence of a colored-Gaussian or non-Gaussian noise. In this paper, an AR detector is proposed and tested for temporally colored Gaussian noise in state estimators to detect random false data injection in the smart grid.

\textbf{State Estimation:} 
The state estimation is first proposed by F.~Schweppe in 1970\cite{schweppe1970power1,schweppe1970power2,schweppe1970power3} as a weighted least-squares (WLS) problem. It is since enriched by numerous studies in the following decades \cite{bose1987real,m1990bibliography,abur2004power}. For a power system with $K$ state variables $\bfth = [\theta_1, \theta_2, \ldots, \theta_K]^T$, we have $M$ meter measurements $\mathbf{x} = [x_1, x_2, \ldots, x_M]^T$, which is given by
\begin{align}
\label{data_model}
\mathbf{x} = \mathbf{H} \bfth + \mathbf{w}
\end{align}
where $\mathbf{H}$ is a $M \times K$ ($M > K$) Jacobian topological matrix and $\mathbf{w}$ is a $M \times 1$ measurement error (noise) vector. The state variables typically include the amplitudes and the phases of voltages in buses. Commonly, the measurement error is modeled by the white Gaussian distribution, i.e., $\mathbf{w} \sim \mathcal{N}(\mathbf{0}, \bfSig_w)$. For this case, it is well-known that the estimated state $\hat{\bfth}$ can be given by the same solution using maximum likelihood estimation (MLE) and weighted least squares (WLS) \cite{Kay:1993_estimation}, as follows:
\begin{align}
\label{no_attack_solution}
\hat{\bfth} = (\mathbf{H}^T \bfSig_w^{-1} \mathbf{H} )^{-1} \mathbf{H}^{T} \bfSig_w^{-1} \mathbf{x}
\end{align}

\textbf{Detection of False Data Injection:} False data injection has been recently identified as a critical type of malicious data attacks in a power system \cite{kosut2011malicious, srivastava2013modeling, yuan2011modeling}. It is necessary to detect the false data injection to protect the safety and the integrity of the power system. Technically, the false data injection can be modeled as follows:
\begin{align}
\label{data_model_attack}
\mathbf{x} = \mathbf{H} \bfth + \mathbf{a} + \mathbf{w}
\end{align}
where $\mathbf{a}$ is the false data injected to the measurements. $\mathbf{a}$ is usually a sparse vector due to the fact that the attacker can only get access to a limited number of component measurements in power system. We call a vector $\mathbf{a}$ has sparsity of $d$ if there are at most $D$ non-zero elements in the vector, i.e., $\| \mathbf{a} \|_{0} = D$.  For power engineers, both the state variables $\bfth$ and the false data $\mathbf{a}$ in Eq. (\ref{data_model_attack}) are unknown. Note that the false data with the form of $\mathbf{a} = \mathbf{H} \bfth_a$ cannot be detected without knowing prior knowledge of state variables, commonly termed unobservable false data. Attackers may fabricate such attacks by gaining intelligence on the system topology $\mathbf{H}$.
Mathematically, the following theorem further indicates whether a false data is observable:
\newtheorem{theorem}{Theorem}\begin{theorem}
\textit{For a given measurement matrix $\mathbf{H}$ with a size of $M \times K$, where $M > K$, there always exists a $M \times (M -K)$ matrix $\mathbf{B}$ such that the columns of $\mathbf{B}$ span the orthogonal subspace of the columns of $\mathbf{H}$, i.e., $\mathbf{B}^T \mathbf{H} = \mathbf{0}$ and $\mathbf{B}^T \mathbf{B} = \mathbf{I}$. Then any false data $\mathbf{a} \in \mathcal{R}^{M}$ can be written as: $\mathbf{a} = \mathbf{H} \bfth_a + \mathbf{B} \bfth_b$, where $\bfth_a \in \mathcal{R}^{K}$ and $\bfth_b \in \mathcal{R}^{M-K}$.}
\end{theorem}

The proof of Theorem 1 follows the Orthogonal Decomposition Theorem \cite{adkins2012algebra} directly. From Theorem 1, it can be shown that the false data is observable when $\bfth_b \neq \mathbf{0}$, and $\bfth_a$ is part of state which is estimated as $\hat{\bfth} = \bfth + \bfth_a$. Defining a new state variable as $\bfth_1 = \bfth + \bfth_a$, the measurements can be further written as: $\mathbf{x} = \mathbf{H}\bfth_1 + \mathbf{B} \bfth_b + \mathbf{w}$, and the hypothesis testing problem for the false data injection detection becomes: 
\begin{align}
\label{new_hypo}
& \mathcal{H}_0: \bfth_b = \mathbf{0} \nonumber \\
& \mathcal{H}_1: \bfth_b \neq \mathbf{0}
\end{align}

For the measurements corrupted by the noise $\mathbf{w} \sim \mathcal{N}(\mathbf{0}, \mathbf{I})$, we give the following GLRT detector for the hypothesis testing problem in Eq. (\ref{new_hypo}):
\begin{theorem}
\textit{For the given measurement $\mathbf{x} = \mathbf{H}\bfth_1 + \mathbf{B}\bfth_b + \mathbf{w}$, where both $\bfth_1$ and $\bfth_b$ are unknown, $\mathbf{w} \sim \mathcal{N}(\mathbf{0}, \mathbf{I})$, $\mathbf{B}^T \mathbf{H} = \mathbf{0}$ and $\mathbf{B}^T \mathbf{B} = \mathbf{I}$, the GLRT detector for the hypothesis testing problem in Eq. (\ref{new_hypo}) is to decide $\mathcal{H}_1$ if 
\begin{align}
T(\mathbf{x}) & = 2 \ln \frac{p(\mathbf{x}; \hat{\bfth}_1, \hat{\bfth}_b)}{p(\mathbf{x}; \hat{\bfth}_1, \mathbf{0})} \nonumber \\
& = \mathbf{x}^T P_{\mathbf{H}}^{\perp}\mathbf{x} > \tau
\end{align}
where $\tau$ is the threshold, $\mathbf{P}_{\mathbf{H}}^{\perp} = \mathbf{I} - \mathbf{H}(\mathbf{H}^T \mathbf{H})^{-1} \mathbf{H}^T $, and $\hat{\bfth}_1$ and $\hat{\bfth}_b$ are given by
\begin{align}
\label{MLE_2}
& \hat{\bfth}_1 = (\mathbf{H}^T \mathbf{H} )^{-1} \mathbf{H}^{T} \mathbf{x} \nonumber \\
& \hat{\bfth}_b = \mathbf{B}^{T} \mathbf{x}
\end{align}
} 
\end{theorem}
{The proof of Theorem 2 is provided in the Appendix of this paper.} To apply Theorem 2 for false data injection detection with $\mathbf{w} \sim \mathcal{N}(\mathbf{0}, \bfSig_w)$, we first define a pre-whitened variable $\mathbf{y} = \mathbf{M}\mathbf{x}$ where $\mathbf{M}$ is the known whitening transformation matrix such that $\mathbf{M}^T \mathbf{M} = \bfSig_w^{-1}$, and $\mathbf{H}{'} = \mathbf{M}\mathbf{H}$. We then have $\mathbf{y} = \mathbf{H}{'} \bfth + \mathbf{a}{'} + \mathbf{w}{'}$, where $\mathbf{a}{'} = \mathbf{M}\mathbf{a}$ and $\mathbf{w}{'} = \mathbf{M}\mathbf{w}$. Using the pre-whitening transformation, we have $\mathbf{w}{'} \sim \mathcal{N}(\mathbf{0}, \mathbf{I})$.  According to the Theorem 1, there always exists a matrix $\mathbf{B}{'}$ such that ${\mathbf{B}{'}}^{T}\mathbf{H}{'} = \mathbf{0}$ and ${\mathbf{B}{'}}^T \mathbf{B}{'} = \mathbf{I}$, and $\mathbf{a}{'} = \mathbf{H}{'} \bfth_a + \mathbf{B}{'} \bfth_b$. Thus, we have $\mathbf{y} = \mathbf{H}{'} (\bfth + \bfth_a) + \mathbf{B}{'}\bfth_b + \mathbf{w}{'}$. According to the Theorem 2, hence, we have the following GLRT detector for the false data injection detection:
\begin{align}
\label{GLRT}
T(\mathbf{y}) & = \mathbf{y}^T (\mathbf{I} - \mathbf{H}{'}({\mathbf{H}{'}}^T \mathbf{H}{'})^{-1} {\mathbf{H}{'}}^T) \mathbf{y} \nonumber \\
& = \mathbf{x}^T \mathbf{M}^T (\mathbf{I} - \mathbf{H}{'}({\mathbf{H}{'}}^T \mathbf{H}{'})^{-1} {\mathbf{H}{'}}^T) \mathbf{M} \mathbf{x} \nonumber \\
& = \mathbf{x}^T  (\mathbf{I} - \bfSig_{w}^{-1} \mathbf{H}({\mathbf{H}}^T \bfSig_w^{-1} \mathbf{H})^{-1} {\mathbf{H}}^T) \bfSig_{w}^{-1} \mathbf{x} \nonumber \\
\end{align}

\textbf{Sequential Observations:} 
{In FDI attacks, an adversary usually hacks the meter measurements for a period of time to mislead the decision making in power systems. The Cramer-Rao Lower Bound (CRLB) theorem indicates that using multiple observations leads to a much lower variance of the state estimation. Given $N$ sequential observations, the estimated state in power system can be written as:}
\begin{align}
\label{N_estimation_Gaussian}
\hat{\bfth} =  (\mathbf{H}^T \bfSig_w^{-1} \mathbf{H} )^{-1} \mathbf{H}^{T} \bfSig_w^{-1} \bar{\mathbf{x}}
\end{align}
where $\bar{\mathbf{x}} = \sum_{i=1}^N\mathbf{x}_i / N$ is the mean of $N$ observations, and the GLRT detector for an unknown false data can be given by 
\begin{align}
\label{N_detection_Gaussian}
T(\mathbf{X}) & = \frac{2}{N} \ln \frac{p(\mathbf{X}; \hat{\bfth}_1, \hat{\bfth}_b)}{p(\mathbf{X}; \hat{\bfth}_1, \mathbf{0})} \nonumber \\ 
& =\bar{\mathbf{x}}^T (\mathbf{I} - \bfSig_{w}^{-1} \mathbf{H}({\mathbf{H}}^T \bfSig_w^{-1} \mathbf{H})^{-1} {\mathbf{H}}^T) \bfSig_{w}^{-1} \bar{\mathbf{x}}
\end{align}
{Here, we assume that the false data is a targeted measurement vector that the attacker intends to inject over the $N$ observations. It can be further extended to state-dependent or random but disruptive vectors under complex attack schemes. The variation of states is also negligible during the period of $N$ observations. As a simplification of complex power system dynamics, this steady-state assumption is commonly used in many studies on DC state estimation and contingency analysis \cite{kosut2010malicious}\cite{kosut2010maliciousc}. To further validate this assumption, we provide more simulation results in the Supplemental Material with dynamic loading change in the system.}

\section{Proposed State Estimator and False Data Detector with Colored Gaussian Noise}
We consider the problems of both state estimation and false data detection in power systems when the measurements are corrupted by the colored Gaussian noise which is modeled by an AR process, and the conventional estimator and detector with the Gaussian noise can be considered as the special case of our methodology. The observation matrix $\mathbf{X}$ can be rewritten as $\mathbf{X} = [\mathbf{x}_1, \mathbf{x}_2, \ldots, \mathbf{x}_M]^T$, and the measurement error (noise) matrix $\mathbf{W}$ can be rewritten as $\mathbf{W} = [\mathbf{w}_1, \mathbf{w}_2, \ldots, \mathbf{w}_M]^T$, where both $\mathbf{x}_i$ and $\mathbf{w}_i$ are $N \times 1$ vectors. For a power system without false data injection, we have
\begin{align}
\label{s_vector}
\mathbf{x}_i = \mathbf{1}_N \mathbf{h}_i^T \bfth + \mathbf{w}_i,  \quad i = 1, 2, \ldots, M
\end{align}
where $\mathbf{h}^T_i$ is the $i$-th row of $\mathbf{H} = [\mathbf{h}_1, \mathbf{h}_2, \ldots, \mathbf{h}_M]^T$, and $\mathbf{1}_N$ is a $N \times 1$ all-ones vector. Unlike the existing approaches with the assumption of white Gaussian noise, we consider that the sequential noise $\mathbf{w}_i = [w_{i,0}, w_{i,1}, \ldots, w_{i,N}]^T$ for the $i$-th meter measurement follows a colored Gaussian distribution which is modeled via a $p$-order AR process:
\begin{align}
w_{i, n} = \sum_{j=1}^p \alpha_{i,j} w_{i, n-j} + v_{i,n}, \ n=0, 1, \ldots, N-1
\end{align} 
where $\alpha_{i,1}, \alpha_{i,2}, \ldots, \alpha_{i,p}$ are the parameters of the AR process, and $v_{i,n}$ is an independent and identically distributed (I.I.D.) random variable which satisfies a white Gaussian distribution, i.e., $v_{i,n} \sim \mathcal{N}(0, \sigma_i^2)$. It is known that $w_{i, n}$ and $\mathbf{w}_i$ are also Gaussian. Given $w_{i,-1}, w_{i,-2}, \ldots, w_{i,-p}$, it can be shown that the $N$ sequential observations of the $i$-th measurement have a probability density function (PDF) $p(\mathbf{x}_i; \bfth)$ \cite{kay1983asymptotically}\cite{tang2015parametric} as follows:
\begin{align}
& p(\mathbf{x}_i; \bfth) = \prod_{n=0}^{N-1} p(x_{i,n} | x_{i,n-1}, x_{i,n-2}, \ldots, x_{i,n-p} ) \nonumber \\
& = \frac{1}{(2 \pi \sigma^2_i)^{N/2}} \exp \left\{ -\frac{1}{2 \sigma_i^2} \sum_{n=0}^{N-1} \left[ w_{i,n} - \sum_{j=1}^p \alpha_{i,j} w_{i, n-j}  \right] \right\}
\end{align}
where the exponential term in above equation can be rewritten as
\begin{align}
&\sum_{n=0}^{N-1} \left[ w_{i,n} - \sum_{j=1}^p \alpha_{i,j} w_{i, n-j}  \right] = (\mathbf{T}_i \mathbf{w}_i + \mathbf{c})^T  (\mathbf{T}_i \mathbf{w}_i + \mathbf{c}_i) \nonumber \\
& = (\mathbf{T}_i \mathbf{x}_i + \mathbf{c}_i - \mathbf{T}_i\mathbf{1}_N \mathbf{h}_i^T \bfth)^T (\mathbf{T}_i \mathbf{x}_i + \mathbf{c}_i  - \mathbf{T}_i\mathbf{1}_N \mathbf{h}_i^T \bfth)
\end{align}
where $\mathbf{T}_i$ is given by 

\begin{eqnarray}
\label{T_c_form}
\Scale[0.75]{
\mathbf{T}_i = \left[ \begin{array}{c c c c c c c c c}
&\alpha_{i,0} & 0 & 0 & \cdots & \cdots & \cdots & 0 \\
&-\alpha_{i,1} & \alpha_{i,0} & 0 & \cdots & \cdots & \cdots& 0 \\
 &  & \ddots &   &  & & & \vdots \\
&-\alpha_{i,p} & -\alpha_{i,p-1} & \cdots & \alpha_0 & \cdots & \cdots & 0 \\
&0 & -\alpha_{i,p} & -\alpha_{i,p-1} & \cdots & \alpha_{i,0} &  \cdots & 0 \\
 &  & \ddots &   &  & & & \vdots \\
&0 & \cdots & \cdots & -\alpha_{i,p} & -\alpha_{i,p-1} & \cdots & \alpha_{i,0} \\
\end{array} 
\right]
}
\end{eqnarray}
and $\mathbf{c}_i$ is given by
\begin{eqnarray}
\Scale[0.78]{
\mathbf{c}_i = \left[ -\sum \limits_{k=1}^{p} \alpha_{i,k} x[-k], -\sum \limits_{k=2}^{p}\alpha_{i,k} x[-k], \ldots, -\alpha_{i,p} x[-p], 0, \ldots, 0 \right]^T
}
\end{eqnarray}

Therefore, it can be shown that $\mathbf{T}_i \mathbf{x}_i + \mathbf{c}_i \sim \mathcal{N}(\mathbf{T}_i\mathbf{1}_N \mathbf{h}_i^T \bfth, \sigma_i^2 \mathbf{I})$. For all $M$ measurements, we have the PDF $p(\mathbf{X}; \bfth)$ as follows:
\begin{align}
p(\mathbf{X}; \bfth) = \prod_{i=1}^M p(\mathbf{x}_i; \bfth)
\end{align}
and the log-likelihood $J(\bfth)$ as follows:
\begin{align}
& J(\bfth) = \log p(\mathbf{X}; \bfth) \nonumber \\
& = -\sum\limits_{i=1}^M \frac{1}{2 \sigma_i^2} \Scale[0.85]{(\mathbf{T}_i \mathbf{x}_i + \mathbf{c}_i - \mathbf{T}_i\mathbf{1}_N \mathbf{h}_i^T \bfth)^T (\mathbf{T}_i \mathbf{x}_i + \mathbf{c}_i  - \mathbf{T}_i\mathbf{1}_N \mathbf{h}_i^T \bfth) 
}
\end{align}

\textbf{State Estimation:} For a power system without false data injection, using the maximum likelihood estimation criterion, we have the following the state estimation solution
\begin{align}
\label{MLE_AR}
\hat{\bfth} & = \argmax_{\bfth \in \Theta} J(\bfth) \nonumber \\
& = \Scale[1.05]{ \left(\sum\limits_{i=1}^M \frac{\mathbf{1}^T_N \mathbf{T}_i^T \mathbf{T} \mathbf{1}_N}{\sigma^2_i} \mathbf{h}_i \mathbf{h}_i^T\right)^{-1} \left(\sum\limits_{i=1}^M \frac{\mathbf{1}^T_N \mathbf{T}_i^T (\mathbf{T}_i \mathbf{x}_i + \mathbf{c}_i)}{\sigma_i^2} \mathbf{h}_i \right)}
\end{align}
Let $\mathbf{A} = diag(a_1/\sigma^2_1, a_2/\sigma^2_2, \ldots, a_M/\sigma^2_M)$ be a $M \times M$ diagonal matrix, where $a_i = \mathbf{1}_N^T \mathbf{T}_i^T \mathbf{T} \mathbf{1}_N$ for $i=1,2,\ldots, M$, and a $M \times 1$ vector $\mathbf{z} = [z_1 / \sigma^2_1, z_2/\sigma^2_2, \ldots, z_M / \sigma^2_M]^T$, where $z_i = \mathbf{1}_N^T \mathbf{T}_i^T (\mathbf{T}_i \mathbf{x}_i + \mathbf{c}_i)$ for $i=1,2,\ldots, M$, Eq. (\ref{MLE_AR}) can be rewritten as
\begin{align}
\label{N_estimation_AR}
\hat{\bfth} = (\mathbf{H}^T \mathbf{A} \mathbf{H})^{-1} \mathbf{H}^{T} \mathbf{z}
\end{align}

\textbf{Detection of False Data Injection:} Following the Theorem 1 and 2, we first define a pre-whitened variable $\mathbf{y} = \mathbf{M} \mathbf{x}$, where $\mathbf{M}^T\mathbf{M} = \mathbf{A}$, and define $\mathbf{H}{'} = \mathbf{M} \mathbf{H} = [\mathbf{h}_1{'}, \ldots, \mathbf{h}_M{'}]^T$. Since $\mathbf{A}$ is a diagonal matrix, $\mathbf{M}$ is also a diagonal matrix and $\mathbf{M} = diag(\sqrt{a_1} / \sigma_1, \ldots, \sqrt{a_M} / \sigma_M)$. According to the Theorem 1, the $i$-th measurement with false data injection can be written as $\mathbf{y}_i = \mathbf{1}_N \mathbf{h}_i^{'T}\bfth_1 + \mathbf{1}_N \mathbf{b}_i^{'T} \bfth_b + \mathbf{w}_i{'}$, where $\mathbf{b}_i^{'T} \mathbf{h}_i{'} = 0$ and $\mathbf{b}_i^{'T} \mathbf{b}_i{'} = 1$, and $\mathbf{w}_i{'}$ is still modeled by the AR process as follows:
\begin{align}
w_{i, n}{'} = \sum_{j=1}^p \alpha_{i,j} w{'}_{i, n-j} + v{'}_{i,n},  \quad n=0,1,\ldots, N-1
\end{align}
where $v{'}_{i,n} \sim \mathcal{N}(0, a_i)$. Hence, it can be shown that $\mathbf{T}_i \mathbf{y}_i + \mathbf{c}_i \sim \mathcal{N}(\mathbf{T}_i\mathbf{1}_N \mathbf{h}_i^{'T} \bfth_1 + \mathbf{T}_i\mathbf{1}_N \mathbf{b}_i^{'T} \bfth_b, a_i \mathbf{I})$. The MLEs of $\bfth_1$ and $\bfth_b$ can be given by
\begin{align}
\label{estimation_AR_detection}
\hat{\bfth}_1 & = (\mathbf{H'}^{T} \mathbf{H}{'})^{-1} \mathbf{H'}^{T} \mathbf{z}{'} \\
\hat{\bfth}_b & = \mathbf{B'}^{T} \mathbf{z}{'}
\end{align}
where $\mathbf{z}{'} = [z{'}_1 / a_1, \ldots, z'_M / a_M]^T$ and $z'_i = \mathbf{1}_N^T \mathbf{T}_i^T (\mathbf{T}_i \mathbf{y}_i + \mathbf{c}_i)$ for $i=1,2,\ldots, M$. Hence, for the hypothesis testing problem in Eq. (\ref{new_hypo}), we have the following GLRT detector for the false data injection detection:
\begin{align}
\label{N_detection_AR}
T(\mathbf{Y}) & = \frac{2}{N} \ln \frac{p(\mathbf{Y}; \hat{\bfth}_1, \hat{\bfth}_b)}{p(\mathbf{Y}; \hat{\bfth}_1, \mathbf{0})} \nonumber \\
& = \frac{1}{N} \mathbf{z'}^{T} (\mathbf{I} - \mathbf{H}{'} (\mathbf{H'}^{T} \mathbf{H}{'})^{-1} \mathbf{H'}^{T}) \mathbf{z}{'}
\end{align}

Note that the conventional Gaussian solution can be considered as a special case in our AR solution, since the distribution modeled by the AR($0$) process is just the white Gaussian distribution. It can be verified that, for the AR(0) process, we have $\mathbf{T} = \mathbf{I}$, $\mathbf{M} = \sqrt{N} \bfSig_w^{-1/2}$, $\mathbf{c} = \mathbf{0}$, and $\mathbf{z}{'} =  N \mathbf{M} \bar{\mathbf{x}}$, and thus the conventional Gaussian solution for false data estimation given by Eq. (\ref{N_detection_Gaussian}) is equivalent to our AR solution given by Eq. (\ref{N_detection_AR}). 

\textbf{Detection of Unobservable False Data Injection:}
{The GLRT detector derived from Theorem 1 and Theorem 2 above has an underlying scenario that assumes the topological information (Jacobian matrix $\mathbf{H}$) is known to the operator/detector but unknown to the attacker, and detector's knowledge of both $\mathbf{H}$ and $\mathbf{X}$ gains an advantage against such attacks. Utilizing this advantage, it is also able to detect certain unobservable attack schemes utilizing the knowledge of observation matrix~$\mathbf{X}$. }

{In \cite{esmalifalak2011stealth}\cite{huang2013bad}, the authors proposed an unobservable attack scheme that does not rely on the knowledge $\mathbf{H}$. Instead, the linear independent component analysis (ICA) based scheme only requires a number of sequential measurements at the steady-state, i.e., the observation matrix $\mathbf{X}$, to bypass traditional Gaussian detectors. The idea is to rewrite $\mathbf{H {\bfth}}$ as $\mathbf{HAy}$, where $A$ is the unknown mixing matrix and $\mathbf{y}$ is the source vector of independent latent variables (components). Then, let $\mathbf{G = HA}$, we have $\mathbf{X = H{\bfth} + w = Gy + w}$. In a noise-free scenario, ICA infers both $ \mathbf{G} $ and $ \mathbf{y} $ so that $\mathbf{X = G y} $ with maximal independency/nongaussianity in $ \mathbf{y} $.  $\mathbf{G}$ has the same number of rows as $\mathbf{H}$, and its columns correspond to the estimated independent components in $\mathbf{y}$; the rows of $\mathbf{y}$ contain the independent components and the columns correspond to the $N$ sequential observations. The inferred $ \mathbf{y} $ is also called quasi-state vector.}

{In the context of state estimation, when the system dynamics changes within a small range, $|\mathbf{X - G y}|$ will be sufficiently smaller than a trivial number $ \epsilon $\cite{esmalifalak2011stealth}. In this scenario, even without knowledge of the actual Jacobian matrix $\mathbf{H}$, the attacker can use the \emph{virtual} Jacobian matrix $\mathbf{G}$ to generate false data with $\mathbf{G \delta \mathbf{y}}$, where $\delta \mathbf{y} = \theta_a$ is the false state in Theorem 1. Such attacks pose threat to the power system state estimation, and they provide an alternative way when the attackers could not gain access to the entire grid topology but a few snapshots of the system measurements are available.}

{Practically, the FastICA algorithm\cite{hyvarinen1999fast}\cite{hyvarinen2004independent} is used to compute the two matrices $\mathbf{G}$ and $\mathbf{y}$.Given a threshold $\varepsilon$ and an initial weight vector $ \mathbf{w} $ of projection $\mathbf{y = w^T z} $, FastICA maximizes the nongaussianity of $\mathbf{y}$ by iteratively updating $ \mathbf{w} $ and computing $\mathbf{G}$ from $\mathbf{w^T G = I}$, where I is the identity matrix. The number of independent components is initially set to the number of samples and iteratiely reduced if certain component has an eigenvalue smaller than a threshold.}


\section{Simulations and Analysis}
%
%
\subsection{{Detection Performance}}
We conduct numerical simulations on the IEEE 30-bus power system to evaluate the detection performance for false data injection. We use MATPOWER, a Matlab package for power system simulation \cite{zimmerman2011matpower}, to extract the measurement matrix $\mathbf{H}$ which has the size of $284 \times 60$. There are 60 states in total which are the voltage amplitudes and angles on the 30 buses, and 284 meter measurements in all buses and branches. For the simplicity of our simulations, we simulate that the noises of all measurements in both observable and unobservable attacks have the same AR process: $e_{i,n} = 0.9 \times e_{i, n-1} + v_{i, n}$, where $v_{i,n} \sim (0, \sigma^2)$ for $i=1,2,\ldots,M$ and $n=1,2,\ldots, N$ ($M=284$ and $N = 20$ are used in our simulations), and the constant false data with the magnitude of $A$ is injected into $D$ random meter measurements (i.e., $\|\mathbf{a} \|_0 = D$). All simulation results reported in this section are averaged over $10,000$ independent runs.

\begin{figure*}
\captionsetup[subfigure]{labelformat=empty}
  \centering
  \subfloat[]{\includegraphics[width=0.33\textwidth]{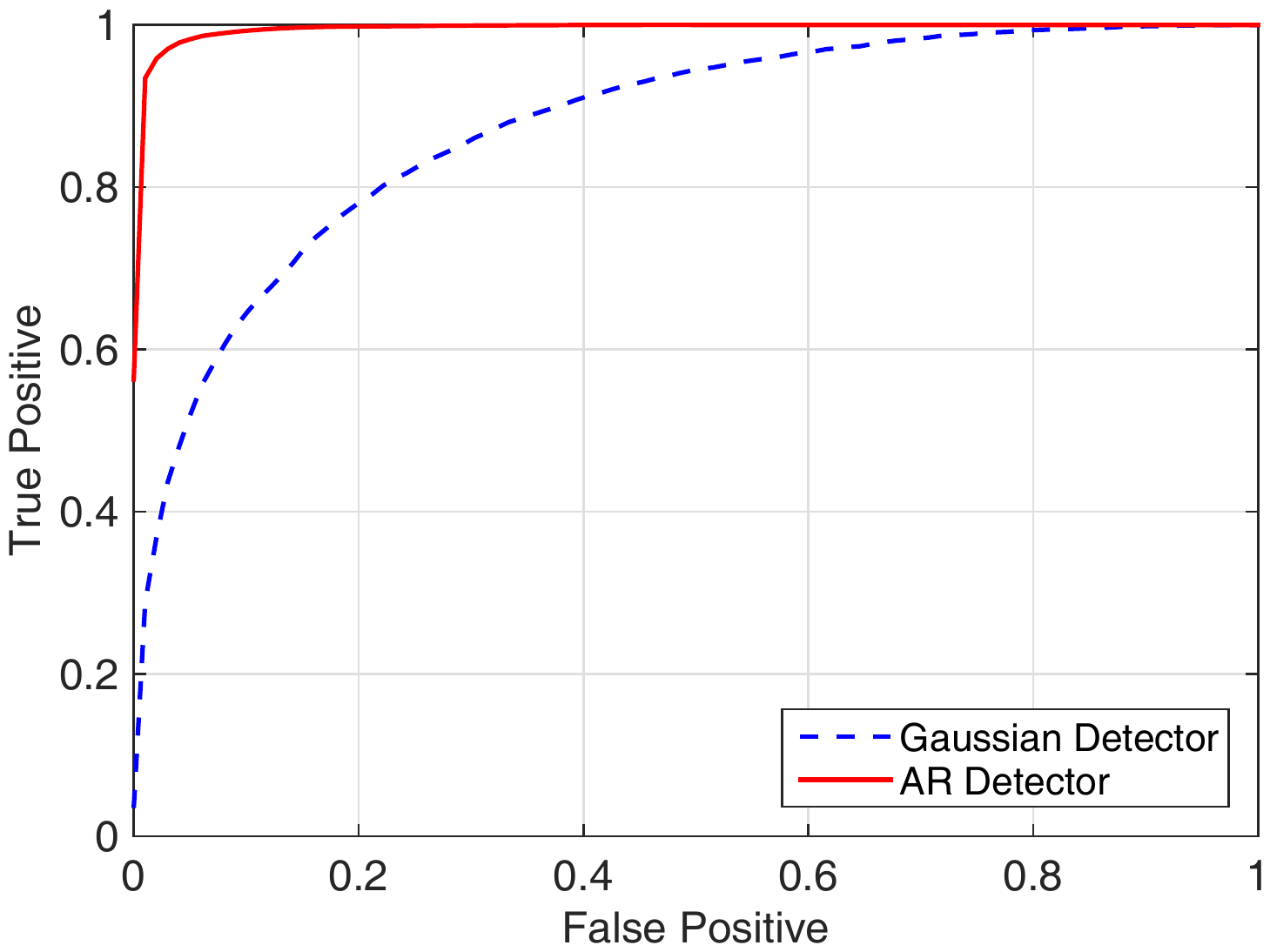}\label{roc_sigma_p3_L10}}
  \subfloat[]{\includegraphics[width=0.33\textwidth]{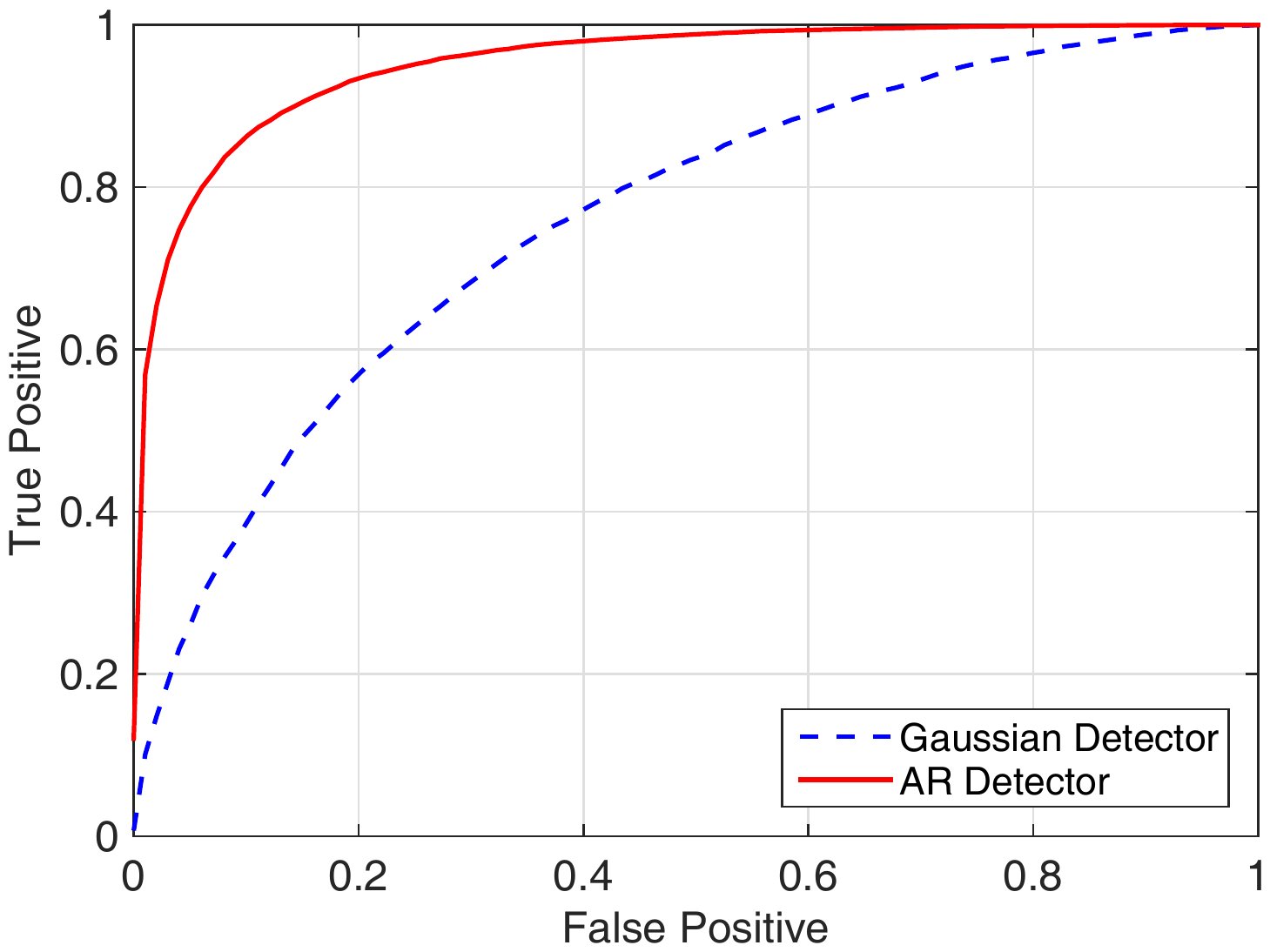}\label{roc_sigma_p5_L10}}
  \subfloat[]{\includegraphics[width=0.33\textwidth]{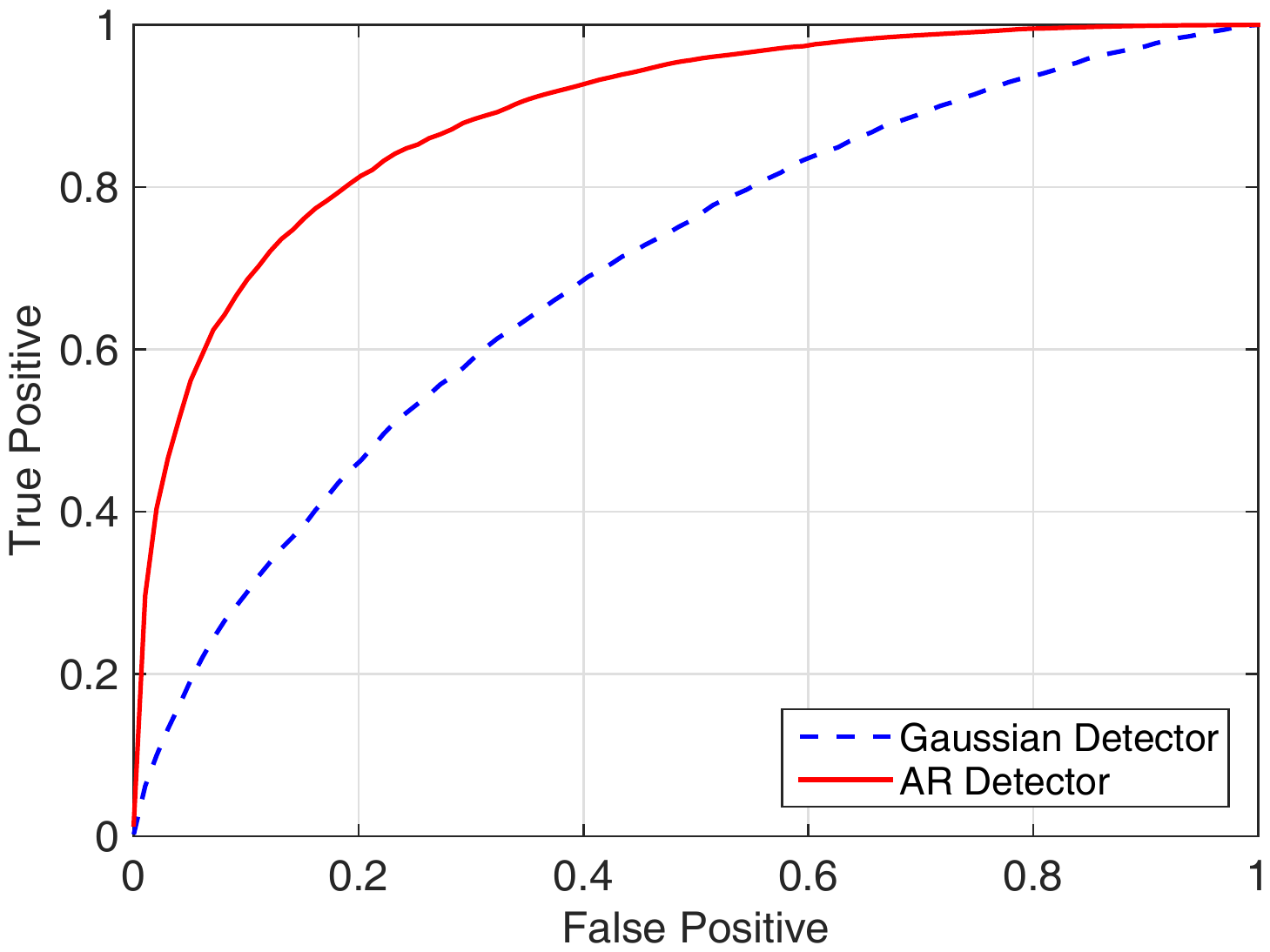}\label{roc_sigma_p7_L10}}  
  \caption{Comparison of ROC curves in observable attacks when: (a) $\sigma^2 = 0.3$, $A=1$, and $D = 29$; (b) $\sigma^2 = 0.5$, $A=1$, and $D = 29$; (c) $\sigma^2 = 0.7$, $A=1$, and $D = 29$}
  \label{roc_sigma_p357}
\end{figure*}
\begin{figure*}
\captionsetup[subfigure]{labelformat=empty}
  \centering
  \subfloat[]{\includegraphics[width=0.33\textwidth]{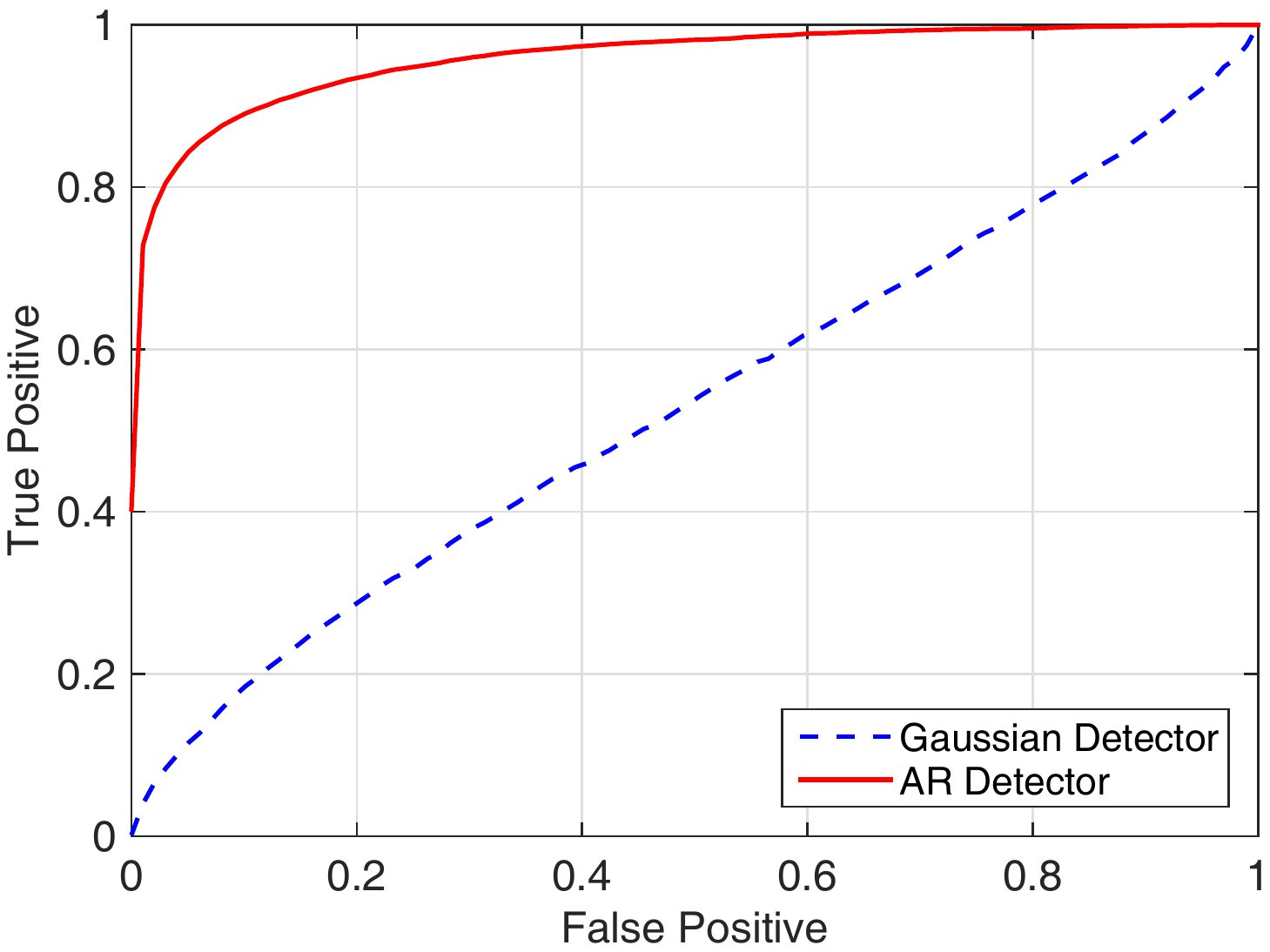}\label{roc_sigma_p3_ICA030}}
  \subfloat[]{\includegraphics[width=0.33\textwidth]{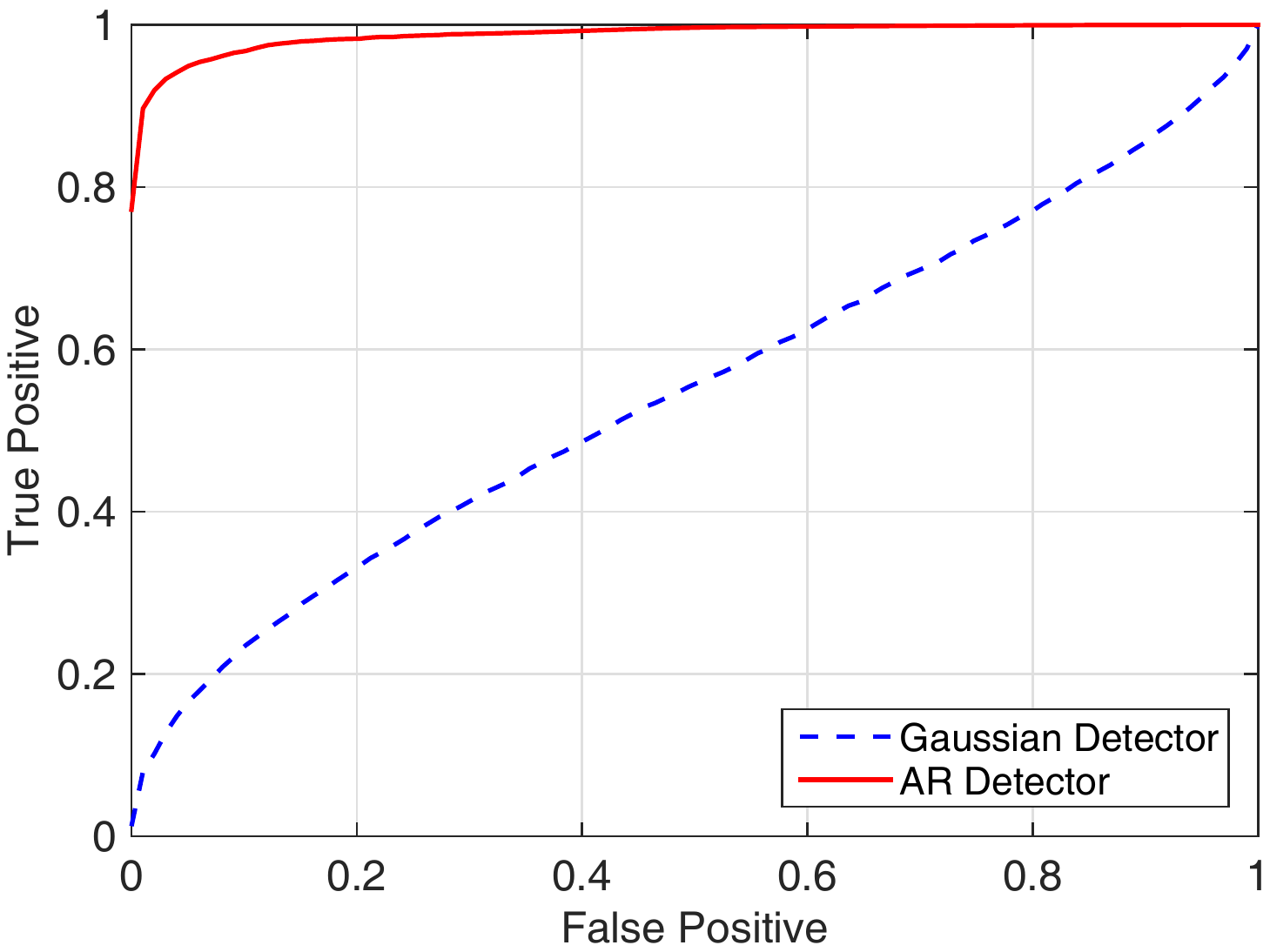}\label{roc_sigma_p3_ICA050}}
  \subfloat[]{\includegraphics[width=0.33\textwidth]{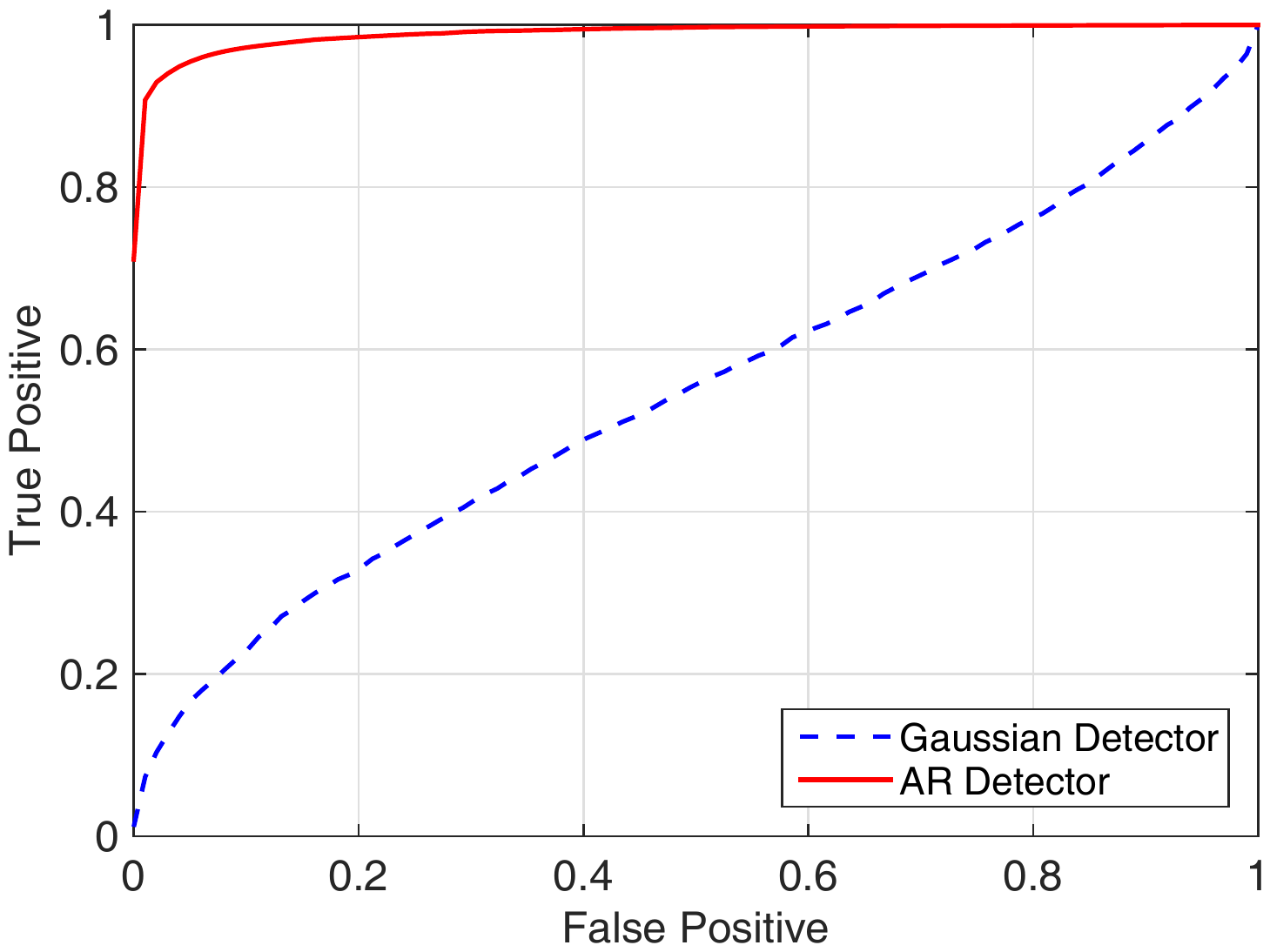}\label{roc_sigma_p7_ICA050}}
  \caption{Comparison of ROC curves in ICA-based unobservable attacks when: (a) $\sigma^2 = 0.3, \sigma^2_y = 0.3$, $A=1$; (b) $\sigma^2 = 0.3, \sigma^2_y = 0.5$, $A=1$; (c) $\sigma^2 = 0.7$, $\sigma^2_y = 0.5$, $A=1$.}
  \label{roc_sigma_ICA}
\end{figure*}
For the observable attacks, we consider that $10\%$ meter measurements (i.e., $D = 29$) can be manipulated by the attackers and that the magnitude of the constant false data is fixed at $A = 1$. Fig.~\ref{roc_sigma_p357} shows the receiver operating characteristic (ROC) curves of the AR detector and the Gaussian detector when { $\sigma^2 = 0.3$, $0.5$ and $0.7$}. It can be shown that the performance of the Gaussian detector is degraded when the assumption of Gaussian noise is not satisfied. The superior performance of the AR detector further demonstrates the effectiveness of the proposed approaches. 



{For the ICA-based unobservable attacks, we consider all measurements are attackable and compare  different combinations of $\sigma^2$ and $\sigma^2_y$ in three cases. Using the same setup for other parameters, the detection performances of unobservable false data attack are shown in Fig.~\ref{roc_sigma_ICA}. The AR detector shows competitive performance over the Gaussian detector against different $\sigma^2_y$ with a given $\sigma^2$ (Fig.~\ref{roc_sigma_p3_ICA030} and Fig.~\ref{roc_sigma_p3_ICA050}) and remains robust against different $\sigma^2$ with a given $\sigma^2_y$ (Fig.~\ref{roc_sigma_p3_ICA050} and Fig.~\ref{roc_sigma_p7_ICA050}). The performance of Gaussian detector deteriorates significantly in unobservable attacks when the inference of virtual Jacobian matrix and quasi-state vector by the ICA based scheme. }



\subsection{{Robustness Analysis}}
{From both the attack and defense perspective, the power system is assume to hold a steady-state for the duration of $N$ samples. With $N = 20$ in a PMU-equipped power system, this duration will be less than half a second at a rate of 48 samples per second. In practice, this assumption usually holds when the system dynamics remain within a certain range.}

In the following simulations, we verify that the assumption of constant states for a short period of time does not reduce the performance of state estimation (SE) and false data injection (FDI) detection, if the system loading changes within a small range. This assumption commonly holds in the study of FDI [19]. More specifically, we show that the performance difference of both SE and FDI methods for a constant full system load and a dynamic system load is small enough to be negligible. 

\begin{figure}
\centering
\includegraphics[width = 0.49\textwidth]{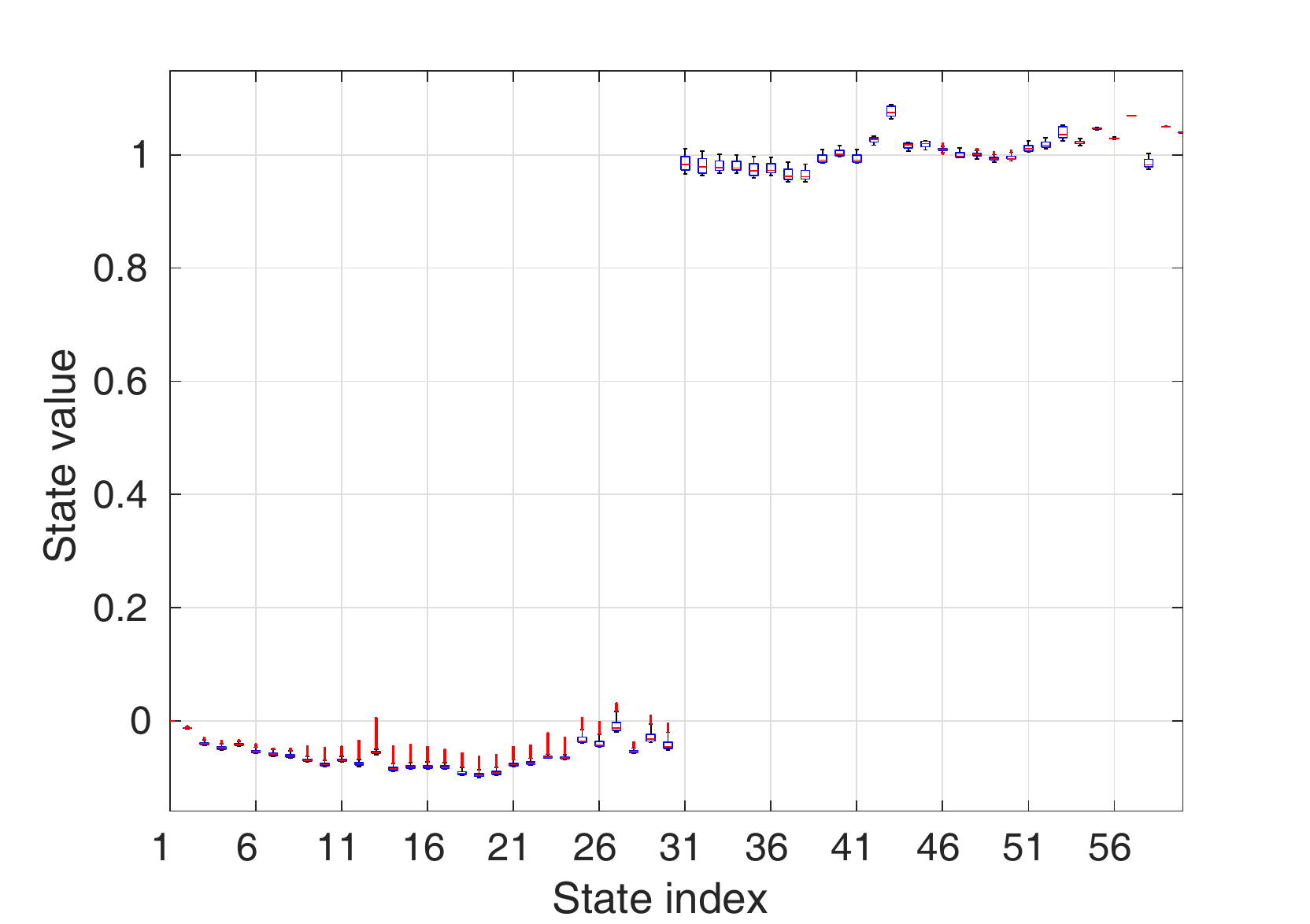}
\caption{The variations of 60 states when the system loading ranges from $95\%$ to $105\%$ for the IEEE 30-bus power system.} 
\label{state_variation}
\end{figure}
\subsubsection{State Variation with Dynamic Load}
\begin{figure*}
  \centering
  \subfloat[]{\includegraphics[width=0.42\textwidth]{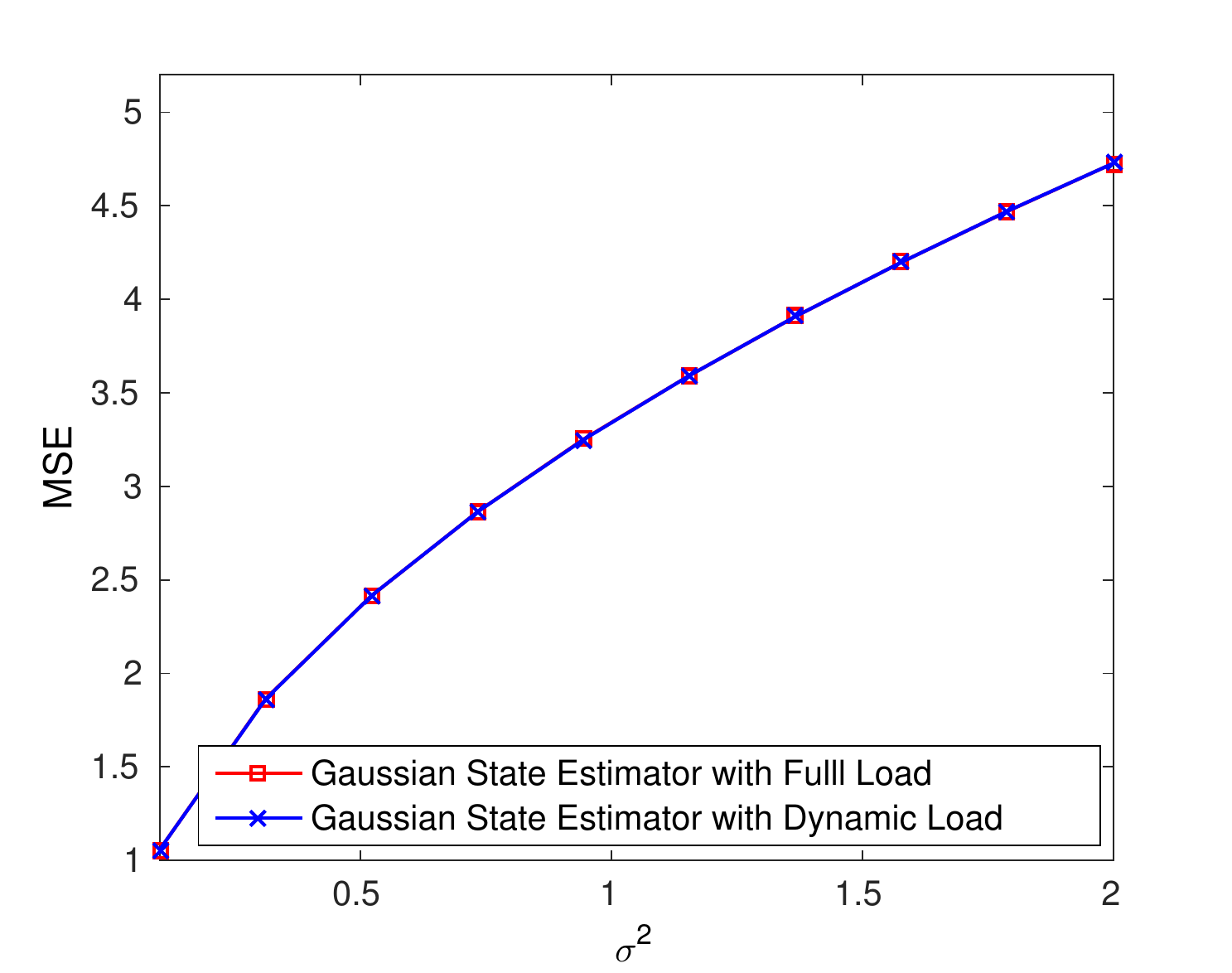}\label{mse_gaussian}}
  \subfloat[]{\includegraphics[width=0.4\textwidth]{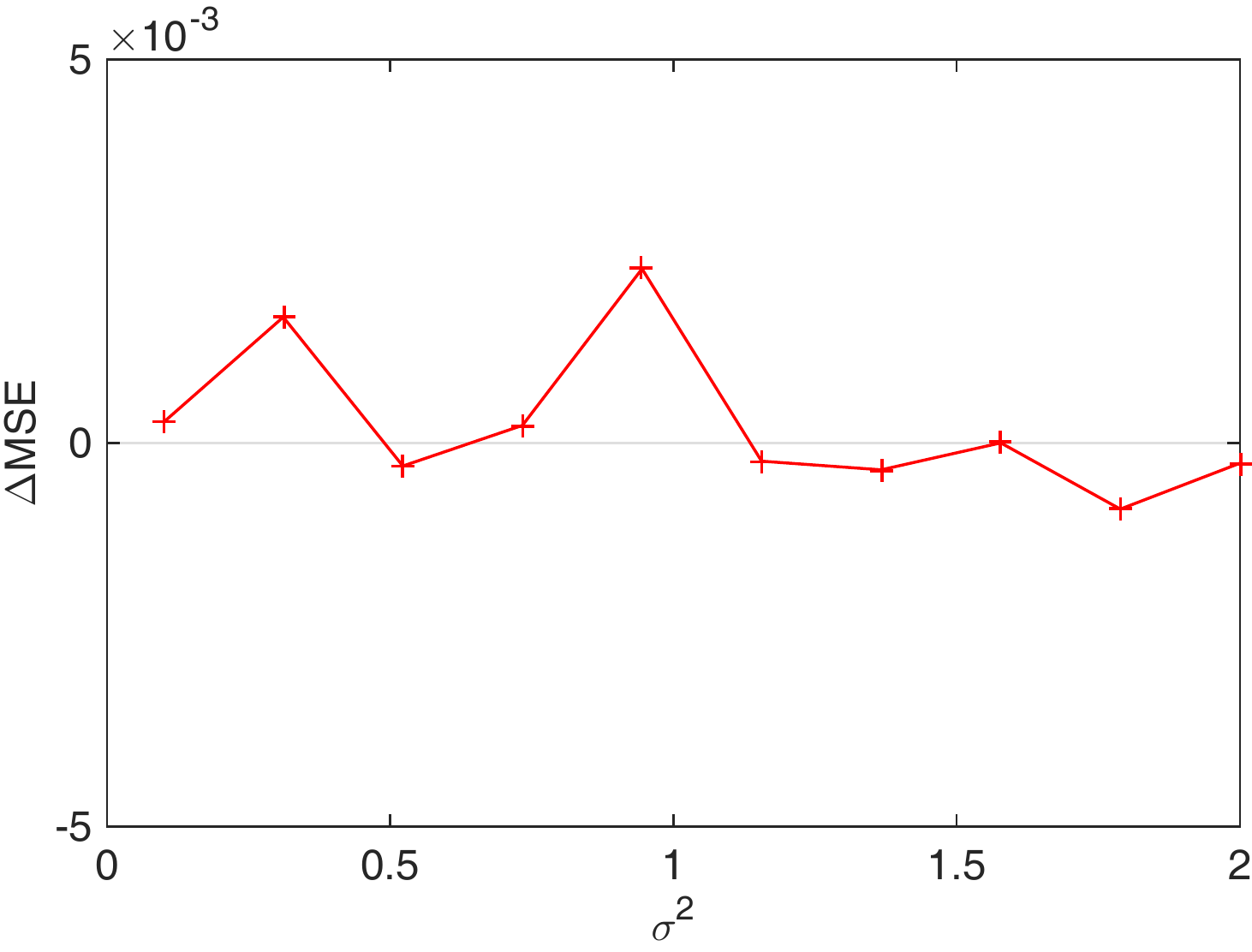}\label{mse_gaussian_diff}}%
  \caption{(a). SE performance variations for two operating points, where the dynamic system loading ranges from $95\%$ to $105\%$; (b). The difference between two MSEs.}
  \label{mse_gaussian_all}
\end{figure*}
We first perform computational simulations to illustrate the small variation of state variables with the change of a power system's load. The states of an AC power system, including the amplitude and the phase of voltages in buses, are determined by the system loading which is usually varied in a small range in practice for a short period of time. For the IEEE 30-bus power system, we obtain 100 operating points whose overall load demand are between $95\%$ to $105\%$ of the original benchmark. The accurate states are then calculated by AC-OPF. We show the variation of states with the dynamic loading in the boxplot of Fig. \ref{state_variation}. The state variation is small in the dynamic load. 

Next, we consider $N$ sequential meter measurements that are corrupted with white Gaussian noise and apply the least square state estimator of the Gaussian approach given by Eq. (8) in our manuscript to estimate the states for two operating points: one is the original base case with $100\%$ load, and another dynamic loading case with a random total load between $95\%$ to $105\%$ of the base case. The results of all rest simulations are averaged over 10,000 runs. In each run, we obtain $N = 20$ sequential observations: $\mathbf{x}_i, i = 1,2,\cdots,20$ where $\mathbf{x}_i$ is a meter measurement vector at $i$-th time frame. Given these sequential observations of meter measurements, we apply Eq. (8) in our manuscript to estimate the state $\hat{\bfth}$, and calculate the MSE of meter measurements as follows:
\begin{align}
MSE = \frac{1}{N}\sum_{i=1}^N \| \mathbf{x}_i - \mathbf{H} \hat{\bfth} \| 
\end{align}

The MSEs of the base case and the dynamic case are shown in Fig. \ref{mse_gaussian_all}, when different variances of Gaussian noise are examined. It shows that the difference of MSEs for these two settings of a power system is small and verifies that the performance of state estimation with the Gaussian noise does not reduce by modeling the dynamic load ranging from $95\%$ to $105\%$ as a constant full load. 

\begin{figure}
\centering
\includegraphics[width = 0.48\textwidth]{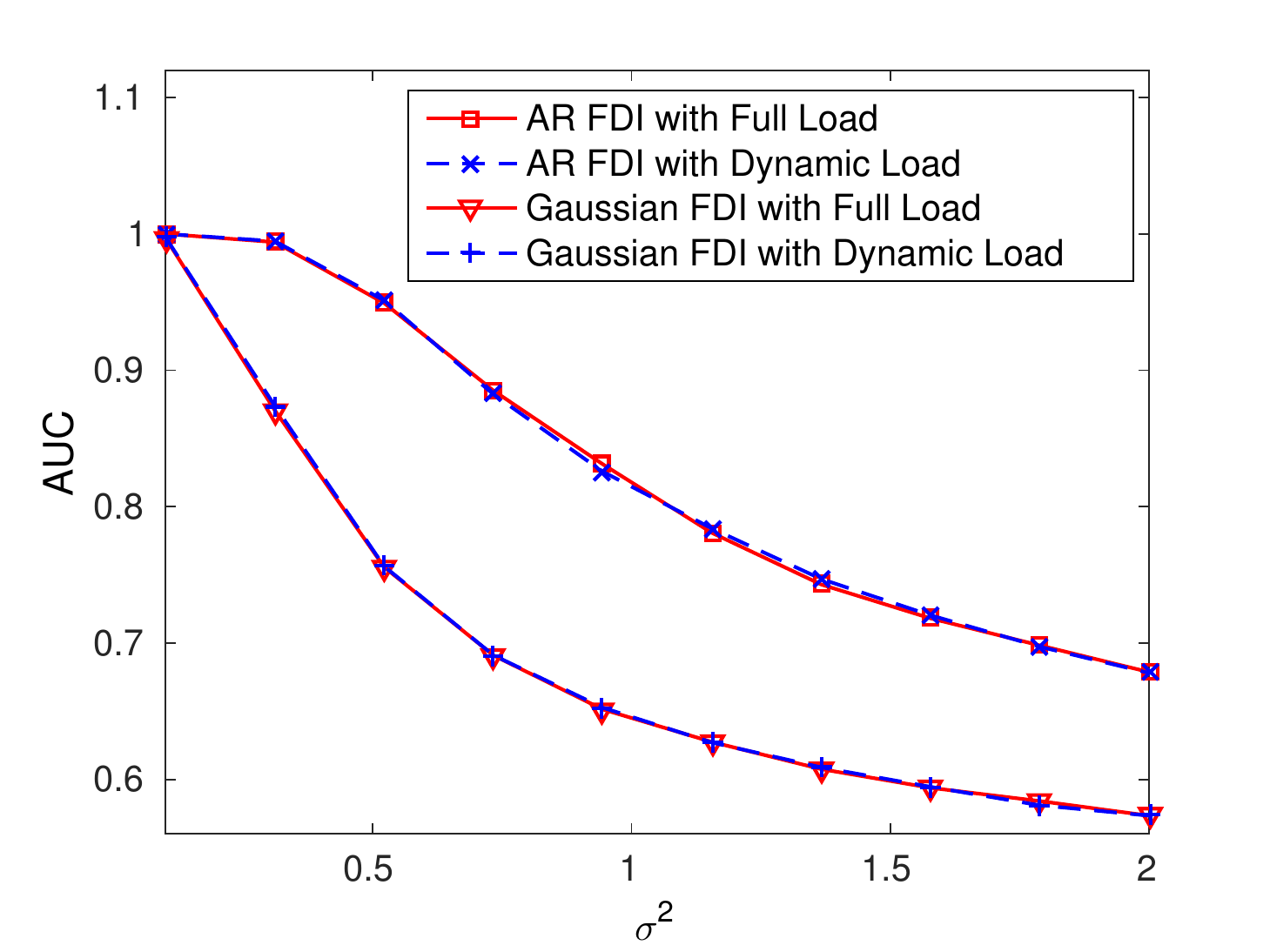}
\caption{FDI detection performance variations and comparisons between the Gaussian and AR detectors, where the dynamic system loading ranges from $95\%$ to $105\%$ for the IEEE 30-bus power system.} 
\label{auc_ar}
\end{figure}

\subsubsection{False Data Injection Detection with Dynamic Load}
We further evaluate the performance variation for FDI detection when the dynamic loading ranges from $95\%$ to $105\%$. We consider the meter measurements are corrupted by the same type of colored Gaussian noise in our manuscript which is modeled as the following AR process: $e_{i,n} = 0.9 \times e_{i, n-1} + v_{i, n}$, where $v_{i,n} \sim (0, \sigma^2)$ for $i=1,2,\ldots,M$ and $n=1,2,\ldots, N$ ($M=284$ and $N = 20$ are used in our simulations). The same false data with the magnitude of $A$ is injected into $D$ random meter measurements (i.e., $\|\mathbf{a} \|_0 = D$). All simulation results are averaged over $10,000$ runs.

For the simulations, we consider that $10\%$ of meter measurements (i.e., $D = 29$) can be manipulated by the attackers and that the magnitude of the constant false data is fixed at $A = 1$. We compare our AR FDI method given by Eq. (22) in our manuscript with the Gaussian FDI method given by Eq. (9) for the base case and dynamic loading case. We show the area under the ROC curve (AUC) performance of these two approaches in Fig. \ref{auc_ar}, when the dynamic system load is considered as a constant full system load. It also shows that the performance variation is small enough to be negligible for both Gaussian and AR FDI detection approaches. 

%
%
%
\section{Conclusion and Future Work}
This paper considered the problems of state estimation and false data detection in power systems, when the measurements are corrupted by the colored Gaussian noise. By modeling the colored Gaussian noise with the autoregressive process, we developed a state estimator and a false data detector to address these two problems. Numerical simulations were performed to demonstrate the effectiveness of the proposed methods. The superior performance of the proposed AR detector demonstrated the potential of AR detector against false data injection attacks in both observable and unobservable cases when the real Jacobian matrix remains secure and confidential to the attackers. In our future work, data-driven machine learning algorithm such as support vector machine \cite{cortes1995support} and nearest neighbor \cite{tang2015enn} will be studied to incorporate the proposed model-driven state estimation methods for FDI detection.

\section*{Acknowledgment}
This work was supported by the National Science Foundation under Grants CNS 1117314 and ECCS 1053717, and by Army Research Office (ARO) under Grant W911NF-12-1-0378.

\appendix[{Proof of Theorem 2}]
\textbf{Theorem 2:}
\textit{For the given measurement $\mathbf{x} = \mathbf{H}\bfth_1 + \mathbf{B}\bfth_b + \mathbf{w}$, where both $\bfth_1$ and $\bfth_b$ are unknown, $\mathbf{w} \sim \mathcal{N}(\mathbf{0}, \mathbf{I})$, $\mathbf{B}^T \mathbf{H} = \mathbf{0}$ and $\mathbf{B}^T \mathbf{B} = \mathbf{I}$, the GLRT detector for the following hypothesis testing problem 
\begin{align}
\label{new_hypoA}
& \mathcal{H}_0: \bfth_b = \mathbf{0} \nonumber \\
& \mathcal{H}_1: \bfth_b \neq \mathbf{0}
\end{align}
is to decide $\mathcal{H}_1$ if 
\begin{align}
T(\mathbf{x}) & = 2 \ln \frac{p(\mathbf{x}; \hat{\bfth}_1, \hat{\bfth}_b)}{p(\mathbf{x}; \hat{\bfth}_1, \mathbf{0})} \nonumber \\
& = \mathbf{x}^T P_{\mathbf{H}}^{\perp}\mathbf{x} > \tau
\end{align}
where $\tau$ is the threshold, $\mathbf{P}_{\mathbf{H}}^{\perp} = \mathbf{I} - \mathbf{H}(\mathbf{H}^T \mathbf{H})^{-1} \mathbf{H}^T $, and $\hat{\bfth}_1$ and $\hat{\bfth}_b$ are given by
\begin{align}
\label{MLE_2A}
& \hat{\bfth}_1 = (\mathbf{H}^T \mathbf{H} )^{-1} \mathbf{H}^{T} \mathbf{x} \nonumber \\
& \hat{\bfth}_b = \mathbf{B}^{T} \mathbf{x}
\end{align}}
\noindent\emph{proof:} Since $\mathbf{w} \sim \mathcal{N}(\mathbf{0}, \mathbf{I})$, we have $\mathbf{x} \sim \mathcal{N}(\mathbf{H}\bfth_1 + \mathbf{B}\bfth_b, \mathbf{I})$. The log-likelihood of of $\mathbf{x}$ is given by
\begin{align}
J(\bfth_1, \bfth_b) = - \frac{1}{2} \left( \mathbf{x} - \mathbf{H}\bfth_1 + \mathbf{B}\bfth_b \right)^T \left( \mathbf{x} - \mathbf{H}\bfth_1 + \mathbf{B}\bfth_b \right) + c
\end{align}
where $c$ is a constant. 

Using the maximum likelihood estimation criterion, both $\bfth_1$ and $\bfth_b$ can be estimated as follows
\begin{align}
& \frac{\partial J(\bfth_1, \bfth_b) }{\partial \bfth_1} = 0 \nonumber  \\ 
& \frac{\partial J(\bfth_1, \bfth_b) }{\partial \bfth_b} = 0 
\end{align}
which leads to 
\begin{align}
& \hat{\bfth}_1 = (\mathbf{H}^T \mathbf{H} )^{-1} \mathbf{H}^{T} \mathbf{x} \nonumber \\
& \hat{\bfth}_b = \mathbf{B}^{T} \mathbf{x}
\end{align}

The GLRT detector can be written as
\begin{align}
T(\mathbf{x}) & = 2 \ln \frac{p(\mathbf{x}; \hat{\bfth}_1, \hat{\bfth}_b)}{p(\mathbf{x}; \hat{\bfth}_1, \mathbf{0})} \nonumber \\
& = \mathbf{x}^T  (\mathbf{I} - \mathbf{H}(\mathbf{H}^T \mathbf{H})^{-1} \mathbf{H}^T) \mathbf{x} > \tau
\end{align}
where $\tau$ is the threshold. If $T(\mathbf{x}) > \tau$, then we accept $\mathcal{H}_1$, otherwise we accept $\mathcal{H}_0$.

\bibliographystyle{IEEEtran}
\bibliography{ref}

\end{document}

%% file: Macros.tex
\newcounter{appcount}
\newcommand{\appendicesname}
            {Appendix\ \thechapter  \Alph{appcount}}
\newcommand{\bookappendicesname}
            {Appendix\ \Alph{appcount}}
\newcommand{\chapterappendix}[1]
          {\par\setcounter{section}{0}
           \setcounter{equation}{0}
           \setcounter{table}{0}
           \setcounter{figure}{0}
          \addtocounter{appcount}{1}   \renewcommand{\theequation}{\thechapter\Alph{appcount}.\arabic{equation}}
          \renewcommand{\thetable}{\thechapter\Alph{appcount}.\arabic{table}}
          \renewcommand{\thefigure}{\thechapter\Alph{appcount}.\arabic{figure}}
           \setcounter{section}{\arabic{chapter}\Alph{section}}
           \if@openright\cleardoublepage\else\clearpage\fi
           \chapter*{\huge{\appendicesname}\newline\newline \Huge{#1}}
           \addcontentsline{toc}{section}{\thechapter\Alph{appcount} #1}
           \markright{\MakeUppercase{\appendicesname.\ { #1}}}}
\newcommand{\bookappendix}[1]
          {\par\setcounter{section}{0}
           \setcounter{equation}{0}
           \setcounter{table}{0}
           \setcounter{figure}{0}
          \addtocounter{appcount}{1}   \renewcommand{\theequation}{\Alph{appcount}.\arabic{equation}}
           \renewcommand{\thetable}{\Alph{appcount}.\arabic{table}}
             \renewcommand{\thefigure}{\Alph{appcount}.\arabic{figure}}
           \setcounter{section}{\arabic{chapter}\Alph{section}}
           \if@openright\cleardoublepage\else\clearpage\fi
           \chapter*{\huge{\bookappendicesname}\newline\newline \Huge{#1}}
           \addcontentsline{toc}{chapter}{\bookappendicesname #1}
          \markright{\MakeUppercase{\bookappendicesname.\ { #1}}} }
\newcounter{example}
\newcounter{property}
\newcommand{\ben}{\begin{equation}}
\newcommand{\een}{\end{equation}}
\newcommand{\bea}{\begin{eqnarray*}}
\newcommand{\eea}{\end{eqnarray*}}
\newcommand{\bean}{\begin{eqnarray}}
\newcommand{\eean}{\end{eqnarray}}

\newcommand{\bfSig}{{\bf \Sigma}}
\newcommand{\bfth}{\mbox{\boldmath{$\theta$}}}

%% file: FalseData_ColoredGaussian.bbl
\begin{thebibliography}{10}
\providecommand{\url}[1]{#1}
\csname url@samestyle\endcsname
\providecommand{\newblock}{\relax}
\providecommand{\bibinfo}[2]{#2}
\providecommand{\BIBentrySTDinterwordspacing}{\spaceskip=0pt\relax}
\providecommand{\BIBentryALTinterwordstretchfactor}{4}
\providecommand{\BIBentryALTinterwordspacing}{\spaceskip=\fontdimen2\font plus
\BIBentryALTinterwordstretchfactor\fontdimen3\font minus
  \fontdimen4\font\relax}
\providecommand{\BIBforeignlanguage}[2]{{%
\expandafter\ifx\csname l@#1\endcsname\relax
\typeout{** WARNING: IEEEtran.bst: No hyphenation pattern has been}%
\typeout{** loaded for the language `#1'. Using the pattern for}%
\typeout{** the default language instead.}%
\else
\language=\csname l@#1\endcsname
\fi
#2}}
\providecommand{\BIBdecl}{\relax}
\BIBdecl

\bibitem{sood2009developing}
V.~Sood, D.~Fischer, J.~Eklund, and T.~Brown, ``Developing a communication
  infrastructure for the smart grid,'' in \emph{IEEE Electrical Power \& Energy
  Conference (EPEC)}, 2009, pp. 1--7.

\bibitem{monticelli2000electric}
A.~Monticelli, ``Electric power system state estimation,'' \emph{Proceedings of
  the IEEE}, vol.~88, no.~2, pp. 262--282, 2000.

\bibitem{sandberg2010security}
H.~Sandberg, A.~Teixeira, and K.~H. Johansson, ``On security indices for state
  estimators in power networks,'' in \emph{Workshop on Secure Control Systems},
  Stockholm, Sweden, 2010.

\bibitem{hug2012vulnerability}
G.~Hug and J.~A. Giampapa, ``Vulnerability assessment of ac state estimation
  with respect to false data injection cyber-attacks,'' \emph{IEEE Transactions
  on Smart Grid}, vol.~3, no.~3, pp. 1362--1370, 2012.

\bibitem{giani2013smart}
A.~Giani, E.~Bitar, M.~Garcia, M.~McQueen, P.~Khargonekar, and K.~Poolla,
  ``Smart grid data integrity attacks,'' \emph{IEEE Transactions on Smart
  Grid}, vol.~4, no.~3, pp. 1244--1253, 2013.

\bibitem{tan2015integrity}
R.~Tan, V.~B. Krishna, D.~K. Yau, and Z.~Kalbarczyk, ``Integrity attacks on
  real-time pricing in electric power grids,'' \emph{ACM Transactions on
  Information and System Security (TISSEC)}, vol.~18, no.~2, p.~5, 2015.

\bibitem{yu2015blind}
Z.-H. Yu and W.-L. Chin, ``Blind false data injection attack using pca
  approximation method in smart grid,'' \emph{IEEE Transactions on Smart Grid},
  vol.~6, no.~3, pp. 1219--1226, 2015.

\bibitem{Cui2012}
S.~Cui, Z.~Han, S.~Kar, T.~Kim, H.~Poor, and A.~Tajer, ``Coordinated
  data-injection attack and detection in the smart grid: A detailed look at
  enriching detection solutions,'' \emph{IEEE Signal Processing Magazine},
  vol.~29, no.~5, pp. 106--115, 2012.

\bibitem{liu2011false}
Y.~Liu, P.~Ning, and M.~K. Reiter, ``False data injection attacks against state
  estimation in electric power grids,'' \emph{ACM Transactions on Information
  and System Security (TISSEC)}, vol.~14, no.~1, p.~13, 2011.

\bibitem{teixeira2010cyber}
A.~Teixeira, S.~Amin, H.~Sandberg, K.~H. Johansson, and S.~S. Sastry, ``Cyber
  security analysis of state estimators in electric power systems,'' in
  \emph{IEEE Conference on Decision and Control}, Atlanta, GA, 2010, pp.
  5991--5998.

\bibitem{plataniotis1997nonlinear}
K.~N. Plataniotis, D.~Androutsos, and A.~N. Venetsanopoulos, ``Nonlinear
  filtering of non-gaussian noise,'' \emph{Journal of Intelligent and Robotic
  Systems}, vol.~19, no.~2, pp. 207--231, 1997.

\bibitem{pitas2013nonlinear}
I.~Pitas and A.~N. Venetsanopoulos, \emph{Nonlinear digital filters: principles
  and applications}.\hskip 1em plus 0.5em minus 0.4em\relax Springer Science \&
  Business Media, 2013, vol.~84.

\bibitem{xu2011state}
W.~Xu, M.~Wang, and A.~Tang, ``On state estimation with bad data detection,''
  in \emph{IEEE Conference on Decision and Control and European Control
  Conference}, Orlando, FL, 2011, pp. 5989--5994.

\bibitem{hagh2011improving}
M.~T. Hagh, S.~M. Mahaei, and K.~Zare, ``Improving bad data detection in state
  estimation of power systems,'' \emph{International Journal of Electrical and
  Computer Engineering (IJECE)}, vol.~1, no.~2, pp. 85--92, 2011.

\bibitem{anwar2014vulnerabilities}
A.~Anwar and A.~N. Mahmood, ``Vulnerabilities of smart grid state estimation
  against false data injection attack,'' in \emph{Renewable Energy
  Integration}, 2014, pp. 411--428.

\bibitem{dan2010stealth}
G.~D{\'a}n and H.~Sandberg, ``Stealth attacks and protection schemes for state
  estimators in power systems,'' in \emph{IEEE International Conference on
  Smart Grid Communications}, Gaithersburg, MD, pp. 214--219.

\bibitem{Liyan2015}
L.~Jia, R.~Thomas, and L.~Tong, ``Malicious data attack on real-time
  electricity market,'' in \emph{IEEE International Conference on Acoustics,
  Speech and Signal Processing (ICASSP)}, 2011, pp. 5952--5955.

\bibitem{vukovic2012network}
O.~Vukovi{\'c}, K.~C. Sou, G.~D{\'a}n, and H.~Sandberg, ``Network-aware
  mitigation of data integrity attacks on power system state estimation,''
  \emph{IEEE Journal on Selected Areas in Communications}, vol.~30, no.~6, pp.
  1108--1118, 2012.

\bibitem{huang2013bad}
Y.~Huang, M.~Esmalifalak, H.~Nguyen, R.~Zheng, Z.~Han, H.~Li, and L.~Song,
  ``Bad data injection in smart grid: attack and defense mechanisms,''
  \emph{IEEE Communications Magazine}, vol.~51, no.~1, pp. 27--33, 2013.

\bibitem{kim2013topology}
J.~Kim and L.~Tong, ``On topology attack of a smart grid: Undetectable attacks
  and countermeasures,'' \emph{IEEE Journal on Selected Areas in
  Communications}, vol.~31, no.~7, pp. 1294--1305, 2013.

\bibitem{esmalifalak2013detecting}
M.~Esmalifalak, N.~T. Nguyen, R.~Zheng, and Z.~Han, ``Detecting stealthy false
  data injection using machine learning in smart grid,'' in \emph{IEEE
  Conference on Global Communications}, Atlanta, GA, 2013, pp. 808--813.

\bibitem{schweppe1970power1}
F.~Schweppe and J.~Wildes, ``Power system static-state estimation, part i:
  Exact model,'' \emph{IEEE Transactions on Power Apparatus and Systems}, vol.
  PAS-89, no.~1, pp. 120--125, 1970.

\bibitem{schweppe1970power2}
F.~Schweppe and D.~Rom, ``Power system static-state estimation, part ii:
  Approximate model,'' \emph{IEEE Transactions on Power Apparatus and Systems},
  vol. PAS-89, no.~1, pp. 125--130, 1970.

\bibitem{schweppe1970power3}
F.~Schweppe, ``Power system static-state estimation, part iii:
  Implementation,'' \emph{IEEE Transactions on Power Apparatus and Systems},
  vol. PAS-89, no.~1, pp. 130--135, 1970.

\bibitem{bose1987real}
A.~Bose and K.~Clements, ``Real-time modeling of power networks,'' \emph{IEEE
  Proceedings}, vol.~75, no.~12, pp. 1607--1622, 1987.

\bibitem{m1990bibliography}
M.~Filho, A.~Leite~da Silva, and D.~Falcao, ``Bibliography on power system
  state estimation (1968-1989),'' \emph{IEEE Transactions on Power Systems},
  vol.~5, no.~3, pp. 950--961, 1990.

\bibitem{abur2004power}
A.~Abur and A.~G. Exposito, \emph{Power system state estimation: theory and
  implementation}.\hskip 1em plus 0.5em minus 0.4em\relax CRC Press, 2004.

\bibitem{Kay:1993_estimation}
S.~Kay, \emph{Fundamentals of Statistical Signal Processing: Estimation
  Theory}.\hskip 1em plus 0.5em minus 0.4em\relax NJ: Prentice-Hall: Englewood
  Cliffs, 1993.

\bibitem{kosut2011malicious}
O.~Kosut, L.~Jia, R.~J. Thomas, and L.~Tong, ``Malicious data attacks on the
  smart grid,'' \emph{IEEE Transactions on Smart Grid}, vol.~2, no.~4, pp.
  645--658, 2011.

\bibitem{srivastava2013modeling}
A.~Srivastava, T.~Morris, T.~Ernster, C.~Vellaithurai, S.~Pan, and U.~Adhikari,
  ``Modeling cyber-physical vulnerability of the smart grid with incomplete
  information,'' \emph{IEEE Transactions on Smart Grid}, vol.~4, no.~1, pp.
  235--244, 2013.

\bibitem{yuan2011modeling}
Y.~Yuan, Z.~Li, and K.~Ren, ``Modeling load redistribution attacks in power
  systems,'' \emph{IEEE Transactions on Smart Grid}, vol.~2, no.~2, pp.
  382--390, 2011.

\bibitem{adkins2012algebra}
W.~Adkins and S.~Weintraub, \emph{Algebra: an approach via module
  theory}.\hskip 1em plus 0.5em minus 0.4em\relax Springer Science \& Business
  Media, 2012, vol. 136.

\bibitem{kosut2010malicious}
O.~Kosut, L.~Jia, R.~J. Thomas, and L.~Tong, ``Malicious data attacks on smart
  grid state estimation: Attack strategies and countermeasures,'' in \emph{IEEE
  International Conference on Smart Grid Communications}, Gaithersburg, MD,
  2010, pp. 220--225.

\bibitem{kosut2010maliciousc}
------, ``On malicious data attacks on power system state estimation,'' in
  \emph{45th International Universities Power Engineering Conference}, Cardiff,
  UK, 2010, pp. 1--6.

\bibitem{kay1983asymptotically}
S.~Kay, ``Asymptotically optimal detection in unknown colored noise via
  autoregressive modeling,'' \emph{IEEE Transactions on Acoustics, Speech and
  Signal Processing}, vol.~31, no.~4, pp. 927--940, 1983.

\bibitem{tang2015parametric}
B.~Tang, H.~He, Q.~Ding, and S.~Kay, ``A parametric classification rule based
  on the exponentially embedded family,'' \emph{IEEE Transactions on Neural
  Networks and Learning Systems}, vol.~26, no.~2, pp. 367--377, 2015.

\bibitem{esmalifalak2011stealth}
M.~Esmalifalak, H.~Nguyen, R.~Zheng, and Z.~Han, ``Stealth false data injection
  using independent component analysis in smart grid,'' in \emph{Smart Grid
  Communications (SmartGridComm), 2011 IEEE International Conference on}.\hskip
  1em plus 0.5em minus 0.4em\relax IEEE, 2011, pp. 244--248.

\bibitem{hyvarinen1999fast}
A.~Hyv{\"a}rinen, ``Fast and robust fixed-point algorithms for independent
  component analysis,'' \emph{Neural Networks, IEEE Transactions on}, vol.~10,
  no.~3, pp. 626--634, 1999.

\bibitem{hyvarinen2004independent}
A.~Hyv{\"a}rinen, J.~Karhunen, and E.~Oja, \emph{Independent component
  analysis}.\hskip 1em plus 0.5em minus 0.4em\relax John Wiley \& Sons, 2004,
  vol.~46.

\bibitem{zimmerman2011matpower}
R.~D. Zimmerman, C.~E. Murillo-S{\'a}nchez, and R.~J. Thomas, ``{MATPOWER}:
  Steady-state operations, planning, and analysis tools for power systems
  research and education,'' \emph{IEEE Transactions on Power Systems}, vol.~26,
  no.~1, pp. 12--19, 2011.

\bibitem{cortes1995support}
C.~Cortes and V.~Vapnik, ``Support-vector networks,'' \emph{Machine learning},
  vol.~20, no.~3, pp. 273--297, 1995.

\bibitem{tang2015enn}
B.~Tang and H.~He, ``{ENN}: Extended nearest neighbor method for pattern
  recognition [research frontier],'' \emph{IEEE Computational Intelligence
  Magazine}, vol.~10, no.~3, pp. 52--60, 2015.

\end{thebibliography}
